\documentclass[preprint,pra,showpacs,preprintnumbers,amsmath,amssymb,showkeys]{revtex4}
\usepackage[dvips]{graphicx}
\usepackage{dcolumn}% Align table columns on decimal point
\usepackage{bm}% bold math
\usepackage{amsmath, amsthm, amssymb}
\begin{document}
\def\kket{\rangle \mskip -3mu \rangle}
\def\bbra{\langle \mskip -3mu \langle}

\def\ket{\rangle}
\def\bra{\langle}

\def\pard{\partial}

\def\sinh{{\rm sinh}}
\def\sgn{{\rm sgn}}

%Characters
\def\alp{\alpha}
\def\del{\delta}
\def\Del{\Delta}
\def\eps{\epsilon}
\def\gam{\gamma}
\def\sig{\sigma}
\def\kap{\kappa}
\def\lam{\lambda}
\def\ome{\omega}
\def\Ome{\Omega}

\def\th{\theta}
\def\vphi{\varphi}

\def\Gam{\Gamma}
\def\Ome{\Omega}

\def\kav{{\bar k}}
\def\vb{{\bar v}}

\def\abf{{\bf a}}
\def\cbf{{\bf c}}
\def\dbf{{\bf d}}
\def\gbf{{\bf g}}
\def\kbf{{\bf k}}
\def\lbf{{\bf l}}
\def\nbf{{\bf n}}
\def\pbf{{\bf p}}
\def\qbf{{\bf q}}
\def\rbf{{\bf r}}
\def\ubf{{\bf u}}
\def\vbf{{\bf v}}
\def\xbf{{\bf x}}
\def\Cbf{{\bf C}}
\def\Dbf{{\bf D}}
\def\Kbf{{\bf K}}
\def\Pbf{{\bf P}}
\def\Qbf{{\bf Q}}

\def\omet{{\tilde \ome}}
\def\gammat{{\tilde \gamma}}
\def\Ft{{\tilde F}}
\def\ut{{\tilde u}}
\def\bt{{\tilde b}}
\def\vt{{\tilde v}}
\def\xt{{\tilde x}}

\def\ph{{\hat p}}

\def\vt{{\tilde v}}
\def\wt{{\tilde w}}
\def\phit{{\tilde \phi}}
\def\rhot{{\tilde \rho}}
\def\Ft{ {\tilde F}}

\def\Cb{{\bar C}}
\def\Nb{{\bar N}}
\def\Ab{{\bar A}}
\def\Db{{\bar D}}
\def\etab{{\bar \eta}}
\def\gb{{\bar g}}
\def\nb{{\bar n}}
\def\bb{{\bar b}}
\def\Pib{{\bar \Pi}}
\def\rhob{{\bar \rho}}
\def\phib{{\bar \phi}}
\def\psib{{\bar \psi}}
\def\omeb{{\bar \ome}}

\def\Sh{{\hat S}}
\def\Wh{{\hat W}}

\def\SS{I}
\def\psiw{{\xi}}
\def\tI{{g}}

\def\Ep#1{Eq.\ (\ref{#1})}
\def\Eqs#1{Eqs.\ (\ref{#1})}
\def\EQN#1{\label{#1}}

\newcommand{\beqa}{\begin{eqnarray}}
\newcommand{\eeqa}{\end{eqnarray}}

%\preprint{APS/123-QED}

\title{Undecidability, entropy and information loss in computations of classical physical systems}
%Bunching and enhancement of quantum Zeno effect in
% interacting two boson system}% Force line breaks with \\

\author{ Sungyun Kim}
 \email{ksyun@apctp.org}
\affiliation{Asia Pacific Center for theoretical physics, San 31,
Hyoja-dong, Nam-gu, Pohang, Gyoungbuk, Korea, 790-784}

\date{\today}
%\pacs{03.75.Lm, 05.30.Jp, 34.20.Cf} \keywords{two-body system,
%correlation function, few-body dynamics,
%  quantum Zeno, bunching, ultra-cold atoms, atom laser}
\begin{abstract}
We investigate how undecidability enters into computations of
classical physical systems and contributes to the increase of
entropy and loss of information. In actual computation with finite
bit of information capacity we accept inconsistency to avoid
undecidability, which in turn affects entropy of the system. We
apply the Shannon entropy to the discretized Liouvillian system.  It
is shown that for any finite bit of information capacity information
is always lost or the entropy always increases for the probability
density following Hamiltonian dynamics, both in time forward and
time backward direction, thus showing information theoretical
version of second law of thermodynamics. This is due to the
finiteness of information capacity and incompressibility of
probability distribution in Liouville's equation.
\end{abstract}
\maketitle

%\section{\label{sec:level1}First-level heading:\protect\\ The line
%break was forced \lowercase{via} \textbackslash\textbackslash}

 In 1931,  two famous theorems \cite{Goedel} are presented by K. G\"odel. The first theorem states that any
axiomatic system that is strong enough to express natural numbers
contains undecidable statements, which can be neither be proved or
disproved within that system. The second theorem states that no
consistent system
 can be used to prove its own consistency.

After G\"odel's work, interesting variation of these theorems
appeared, notably in computer science and algorithmic information
theory.
 In 1936, A. Turing proved that the halting problem,
 the question of whether or not a Turing machine halts on a given program,
 is undecidable \cite{Turing}. Beginning in late 1960s, G. Chaitin showed that in a formal system with
 $n$ bits of axioms it is impossible to prove that a particular binary
 string is of Kolmogorov complexity greater than $n+c$ \cite{Chaitn}. Put it roughly, he states
 that for arbitrary n bit number it is not possible to make a
 smaller size program which prints that number. It is also shown that majority
 of $n$ bit numbers are maximally complex, i.e. the size of program that can print that
 number is $O(n)$.

 How undecidability enters into physics is also researched and
 undecidable problems in physics are presented \cite{Kanter, Moore}.
 In this article we consider simpler undecidable problems when one
 tries to simulate classical physical system by a computer.
 Many physical systems are expressed with continuous real numbers, but computers
 can deal with only finite bits of information. If one ask to a computer
 "What is the decimal expression of a real number $p$?",
  then the computer tries to answers
 \begin{equation}
   p= x_0 .x_{-1} x_{-2} x_{-3} x_{-4}....., \EQN{decimal}
 \end{equation}
 (in \Ep{decimal}, $x_0 .x_{-1} x_{-2} x_{-3} x_{-4}...$ denotes the decimal
 digits.) but in almost all cases the digits make a non-repeating
 infinite sequence. Computing higher order of digits can be
 extremely hard and whether or not there exists a simple algorithm
 to calculate $p$ is in general undecidable problem, i.e.
 the computer never halts or it
 cannot decide the exact answer in finite time. To solve this problem,
 usually people truncate or round off the answer up to some finite digits and write
 \begin{equation}
 p =x_0.x_{-1} x_{-2} x_{-3} x_{-4}.\EQN{decimaltrun}
  \end{equation}
 By truncation or roundoff, \Ep{decimaltrun} becomes a mathematically false
 statement. The equation becomes inconsistent by avoiding
 undecidability, like G\"odel's two theorems. In numerical simulation this inconsistency is
 called as numerical error or round off error. This inconsistency,
 or roundoff errors,
 in turn contribute the entropy of the result, by reducing
 significant digits of the result.

According to Shannon \cite{Shannon}, the entropy of the discrete
system
  with $n$ event probability $P_1,P_2..., P_n$ is
  defined as
  \beqa
   H \equiv -\sum_{i=1}^n  P_i \log P_i. \EQN{Sentropy}
 \eeqa
From now on we take the base of the logarithm as $2$. This entropy
 is a measure of uncertainty or degree of freedom of the system or
 the information capacity of the system.
 $H$ in \Ep{Sentropy} is always a positive quantity since $0\le P_i
 \le  1$.  When the entropy decreases, we call that the uncertainty
 is reduced or {\it information is gained.}
  For example, if we have an unknown digit $X$ which can be either $0$ or $1$
  with probability $1/2$ each, we have 1 bit of entropy by
  \Ep{Sentropy}. The system has one bit of uncertainty or one bit of
  degree of freedom or the ability to store one bit of information (1 bit information capacity).
 If the unknown digit $X$ is identified
  as $0$, then probability of being $0$ is $1$ and  by \Ep{Sentropy} $H$ becomes zero  bit. The
  uncertainty or degree of freedom is decreased by one bit, and we
  gain 1 bit of information and no degree of freedom in this system.

 Now we consider entropy appearing in the computation of classical physical
 systems. The information loss and entropy increase for chaotic systems
 \cite{schack} and generalized Liouville's systems \cite{Plastino}
are already studied, and as a sources of information loss discarding
of information, interaction with environment or coarse graining are
pointed out.
 In this article we restrict our
 attention to the classical Hamiltonian systems which follow the
 Liouville's equation. We disretize the system and see how the entropy changes with time
 evolution.
 It is shown that the calculation of Shannon entropy for the discretized system
naturally separate the information capacity of the system and
Kullback information, and the information is always same or lost for
any finite discretization of probability density following the
Hamiltonian dynamics.

  Consider the probability distribution
  function $p (\pbf, \qbf,t)$ of a particle in the phase space, which satisfies Liouville's
  equation. Suppose that $p$ has compact support and that support
  is contained by finite size box $\Ome$ with volume $\Ome_0$.
  For numerical computation we discretize the phase space $\Ome$ by $N$
number of uniform box shaped cells, with each cell has the volume
$\Ome_0/N$. Let us denote the cells as $C_1,C_2,..., C_N$, and we
approximate the probability distribution $p (\pbf, \qbf,t)$ inside
the cell $C_i$ as $p_i$, which is the mean value of $p (\pbf,
\qbf,t)$ inside the cell $C_i$. This is the place the small
inconsistency enters. Since we cannot describe the probability
distribution function with infinite precision, we replace the
distribution function $p (\pbf, \qbf,t)$ with a mean value $p_i$
inside the cell (coarse-graining over the cell).

The discretized probability density $p_i$s  satisfy the relation
$\sum_{i=1}^N p_i (\Ome_0/N) =1$ and $0\le p_i \le N/\Ome_0$.
Suppose that
 we have initial condition $p_1^{(0)},p_2^{(0)},...,p_N^{(0)}$
 for every cell $C_i$ at time $t=0$. The probability that the
 particle is in the cell $C_i$ is $p_i^{(0)} (\Ome_0/N)$, and
 the entropy $H^{(0)}$ of the discretized system at $t=0$ is given by
 \begin{eqnarray}
 & &H^{(0)} = - \sum_{i=1}^N p_i^{(0)}  (\Ome_0/N)  \log ( p_i^{(0)}
 (\Ome_0/N)) \nonumber \\
 & &=- \sum_{i=1}^N p_i^{(0)}  (\Ome_0/N)  \log ( p_i^{(0)}
 \Ome_0) - \sum_{i=1}^N p_i^{(0)}  (\Ome_0/N)  \log (1/N) \nonumber
 \\
  & &= - \sum_{i=1}^N p_i^{(0)}  (\Ome_0/N)  \log \bigg ( \frac{p_i^{(0)}}{ (1/
 \Ome_0)}\bigg) +  \log N \EQN{H0}
 \eeqa
 The entropy $H^{(0)}$ has maximum value $\log N$ for uniform distribution, i.e.  when all
 $p_i^{(0)} = 1/ \Ome_0$, and minimum value $0$ when  $p_i^{(0)} = N/
 \Ome_0$ for one specific $i$ and $p_{j\ne i}^{(0)}=0$ for all other
 $j $s. So we have
  \beqa
  & & 0 \le H^{(0)} = - \sum_{i=1}^N p_i^{(0)}  (\Ome_0/N)  \log \bigg ( \frac{p_i^{(0)}}{ (1/
 \Ome_0)}\bigg) +  \log N \le \log N \nonumber \\
  & &- \log N \le - \sum_{i=1}^N p_i^{(0)}  (\Ome_0/N)  \log \bigg ( \frac{p_i^{(0)}}{ (1/
 \Ome_0)}\bigg) \le 0 \EQN{infoEQ}
 \end{eqnarray}
 The meaning of terms in \Ep{H0} are following. The maximum entropy $\log N$ is
 the information capacity or the number of bits allowed for us to describe
 the location of a particle in phase space. For uniform
 distribution, we have no information of the location of the
 particle and all allowed bits are remain unknown. When a particle
 is in the one cell with probability $1$, all unknown bits are fixed
 and uncertainty is $0$. In this case we get maximum information
 within allowed information capacity. The term $- \sum_{i=1}^N p_i^{(0)}  (\Ome_0/N)
 \log  (p_i^{(0)}/{ (1/\Ome_0)})$ in
 relation (\ref{infoEQ}) is called Kullback-Leibler
 divergence \cite{Kullback} or relative entropy with respect to the uniform distribution. We see that this
 term is always non-positive, so this is actually the information of particle location we
 get from the system. Note that this term converges to the integral
 $- \int d\Ome \;  p (\pbf, \qbf,0)
 \log  (p (\pbf, \qbf,0)/{ (1/\Ome_0)})$ as $N$ becomes large, so
 for larger $N$ the information we get is more dependent on the
integrability of $p (\pbf, \qbf,0)$ and less dependent on the number
of discrete cell $N$.

\begin{figure}[htb] % Imported eps example.
\begin{center}
\includegraphics[height=1.6in, width=4.4in]{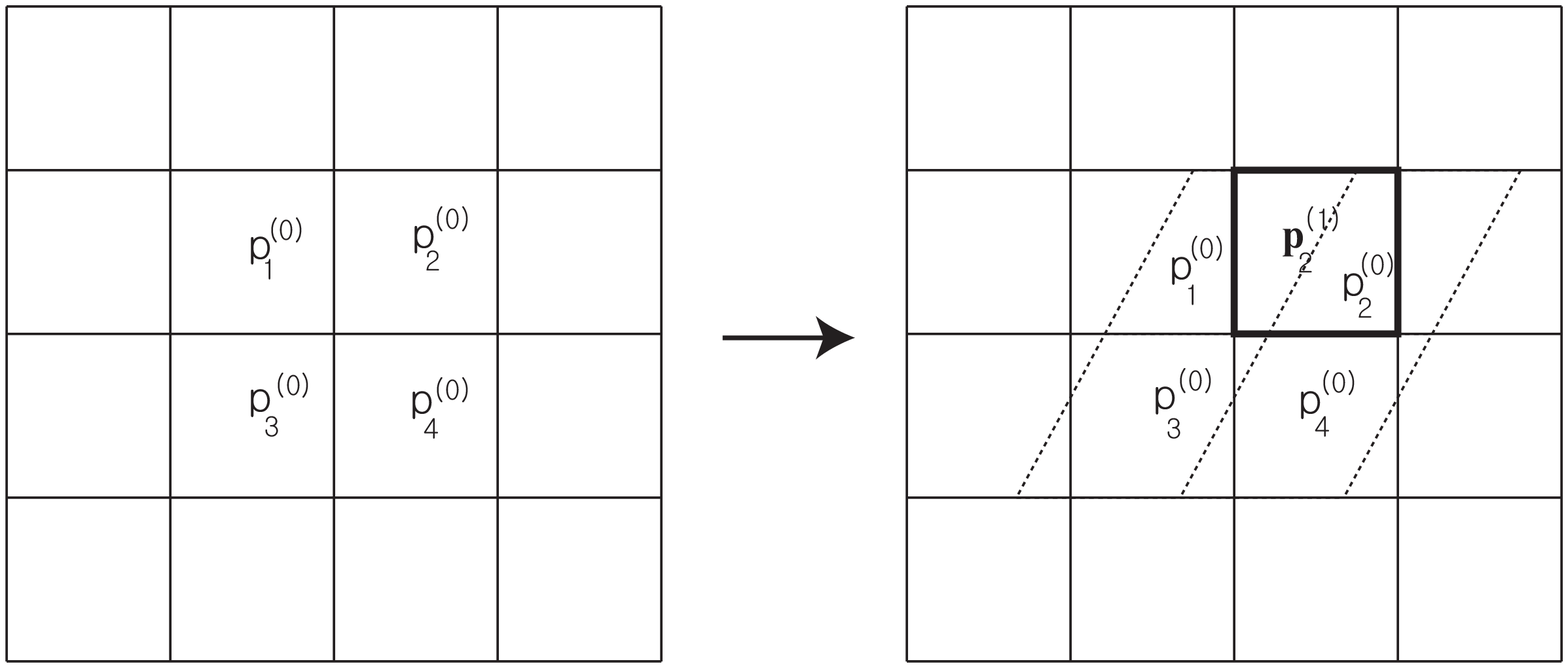}
 \caption{The new discretized probability density $p_i^{(1)}$. In the left figure, each square shaped cells
 has discretized probability density $p_i^{(0)}$s. ($i=1,..,4$) After one discrete time step the cells
 are deformed (shown as dashed parallelograms). The new discretized probability density
 $p_2^{(1)}$ in $C_2$ cell (the square with thick line in the right figure)
 is obtained by averaging the portions of probability densities moved into the $C_2$ cell.} \label{phasecell}
\end{center}
\end{figure}

 Next we consider the time evolution of probability density in this discretized
 system. Since it is a classical Hamiltonian system,
 as time changes the probability density moves like incompressible
 fluid in phase space, i.e. if one follows the time evolution of a point in
 phase space, the density at the representation point remains
 constant and the volume of the neighborhood at the point is conserved.
 The original cell $C_i$s deform, but the
 discretized probability density inside the deformed cell is still $p_i^{(0)}$. Let us denote
 the deformed cells after one discrete time step as $C_i^{(1)}$s. In
 general we cannot track down the deformed cells with
 infinite precision. It is undecidable problem \cite{Moore}.
In practice we see the system with our discretized
 fixed cells of $C_i$s, and
 the new mean discretized probability
 density $p_i^{(1)}$ which is averaged over $C_i$s.  The original cell $C_i$ may
 contain many deformed cell $C_j^{(1)}$s, and
each overlap between $C_j^{(1)}$ with $C_i$ contributes to the
 new mean probability density
 $p_i^{(1)}$  (see figure~\ref{phasecell}). We have
  \begin{eqnarray}
 p_i^{(1)} = \sum_{n=1}^N a_{im} p_m^{(0)} \EQN{Pi1}
 \eeqa
  where $a_{im}$ is given by
  \beqa
  a_{im} = \frac{\mbox{volume of $C_i \cap C_m^{(1)}$}}{\mbox{volume of
  $C_i$}}. \EQN{adef}
  \end{eqnarray}
  This averaging is the place where small inconsistency enters, to
  avoid
  undecidability due to the finite information capacity.
 In \Ep{adef} and from the fact
 $\cup_{m=1}^N C_m^{(1)} = \cup_{i=1}^N C_i = \Ome$ we have the relation
 \begin{equation}
 0 \le a_{im} \le 1, \;\;\;\sum_{m=1}^N a_{im}= \sum_{i=1}^N a_{im}=1. \EQN{aim}
 \end{equation}
  After one time step, the new
 entropy $H^{(1)}$ looking through $C_i$s is
 \begin{eqnarray}
  & &H^{(1)} =- \sum_{i=1}^N p_i^{(1)} (\Ome_0/N) \log \bigg(
  \frac{p_i^{(1)}}
  {1/\Ome_0}\bigg)+ \log N \nonumber \\
  & &=- \sum_{i=1}^N  \sum_{m=1}^N a_{im} p_m^{(0)}  (\Ome_0/N) \log \bigg( \frac{\sum_{m=1}^N a_{im} p_m^{(0)}
  }{
  1/\Ome_0}\bigg)+ \log N. \EQN{H1}
  \end{eqnarray}
 Since the function $f(x) =x \log (\lam x)$ with $\lam>0$ is a convex
 function and the convex function satisfies Jensen's inequality
  \begin{equation}
   f( \sum_i a_i x_i) \le \sum_i a_i f(x_i)\;\;\; \mbox{for all $a_i \ge 0$},\EQN{Jensen}
   \end{equation}
 the first term in RHS of \Ep{H1} is (with
 \Ep{aim})
  \beqa
   & &- \sum_{i=1}^N  \sum_{m=1}^N   (\Ome_0/N) a_{im} p_m^{(0)}  \log
   \bigg( \frac{\sum_{m=1}^N a_{im} p_m^{(0)} }{1/\Ome_0}\bigg) \nonumber \\
  & &\ge - \sum_{i=1}^N \sum_{m=1}^N (\Ome_0/N) a_{im}  p_m^{(0)}
   \log \bigg( \frac{p_m^{(0)} }{1/\Ome_0}\bigg)
   = - \sum_{m=1}^N (\Ome_0/N) p_m^{(0)}
   \log \bigg( \frac{p_m^{(0)}}{1/\Ome_0}
  \bigg). \EQN{entcom}
 \eeqa
 From ( \ref{entcom}) and \Ep{H1} we have
  \beqa
   H^{(0)} \le H^{(1)}, \EQN{H0H1}
   \eeqa
i.e. the information is always lost or the entropy always increases.
Since the property that the probability density distribution moves
like incompressible fluid under Liouville's equation does not change
when it is evolved backward in time, we can do the time evolution of
$P_i^{(0)}$s backward and get the same result $H^{(0)} \le H^{(-1)}$
where $H^{(-1)}$ is the entropy in one discrete time backward. So we
have the information theoretical version of the second law of
thermodynamics, i.e. the information is irreversibly lost during the
time evolution of the classical Hamiltonian system.

 There are two key properties which make the irreversible
 information loss in our case. One is the fixed finite resolution of the
 system, which forces us to take the mean value of the probability density
 over the cell. Second is the incompressibility of the probability distribution
 during time evolution. Without any one of properties the relation
 (\ref{H0H1}) will not hold.

   In conclusion, it is shown how the undecidability enters into
   classical physical system simulation and contribute to the information loss.
   Due to the finiteness of bits we are using, we have to choose
   between undecidability and inconsistency. When we choose
   inconsistency, it affects the uncertainty of the system.
   When we examine the time evolution of the probability distribution in
   the Liouville's equation with finite fixed information capacity, the information is
   always lost in both directions of time or entropy always
   increases. As Jaynes said \cite{Jaynes1,Jaynes2}, this is
   one way of looking statistical mechanics law on the basis of
   the information one can get.

 The author would like to thank Moo Young Choi, Seunghwan Kim and
 Gonzalo Ordonez for helpful comments.

\end{document}